\documentclass[12pt,a4paper]{article}
\usepackage[british]{babel}
\usepackage[a4paper,top=2cm,bottom=2cm,left=2.5cm,right=2.5cm,marginparwidth=1.75cm]{geometry}
\usepackage{cite}
\usepackage{amsmath}
\usepackage{graphicx}
\usepackage[colorlinks=true, allcolors=blue]{hyperref}
\usepackage[title]{appendix}
\usepackage{mathrsfs}
\usepackage{amsfonts}
\usepackage{booktabs}
\usepackage{threeparttable}
\usepackage{algorithm}
\usepackage{algorithmicx}
\usepackage{algpseudocode}
\usepackage{listings}
\usepackage{enumitem}
\usepackage{chngcntr}
\usepackage{booktabs}
\usepackage{lipsum}
\usepackage{authblk}
\usepackage[T1]{fontenc}   
\usepackage{csquotes}      
\usepackage{diagbox}
\usepackage{comment}
\usepackage{subcaption}
\usepackage{caption}
\usepackage{helvet}  
\usepackage{lmodern}  
\usepackage{tabularx}
\usepackage{setspace}
\usepackage{titlesec}
\titleformat{\section} 
  {\normalfont\Large\bfseries}{\thesection.}{1em}{}

\usepackage{lineno} 

\rightlinenumbers 

\linenumbers

\usepackage{float}   
\usepackage{caption} 


\title{Automatic optimisation of a Parallel-Plate Avalanche Counter with Optical Readout}

\author[1, 3]{María Pereira Martínez}
\author[1, 3]{Xabier Cid Vidal}
\author[2, 3]{Pietro Vischia}

\affil[1]{\small Instituto Galego de Física de Altas Enerxías (IGFAE), Universidade de Santiago de Compostela, Santiago de Compostela, Spain}
\affil[2]{\small Universidad de Oviedo and ICTEA, Oviedo, España}
\affil[3]{\small MODE Collaboration \url{https://mode-collaboration.github.io}}

\begin{document}

\maketitle

\begin{abstract}
An automatic optimisation procedure is proposed for some operational parameters of a Parallel-Plate Avalanche Counter with Optical Readout, a detector designed for heavy-ion tracking and imaging. Exploiting differentiable programming and automatic differentiation, we model the reconstruction of the position of impinging 5.5 MeV alpha particles for different detector configurations and build an optimisation cycle that minimizes an objective function. We analyse the performance improvement using this method, exploring the potential of these techniques in detector design.
\end{abstract}

\textbf{Keywords}: Differentiable programming; Machine Learning; Detector optimisation; Nuclear physics.

\section{Introduction}
\label{sec:intro}

Over the last decades, the availability of high-performance computing and the development of deep learning \cite{Goodfellow-et-al-2016} have transformed the optimization of complex systems. When the dimensionality of the space of relevant design parameters exceeds a few units or the relationships between different parameters are not trivial, automated processes can be developed to identify configurations corresponding to the minimum of a carefully specified objective function.

The core of these optimization searches is differentiable programming (DP), a paradigm in which computer programs can be differentiated end-to-end automatically, enabling gradient-based optimization of parameters within the program leveraging automatic differentiation (AD) \cite{diffprogramming, AD_survey, aehle_2022_automatic_forward, aehle_2022_automatic_reverse}. 

Despite the long-established use of AD in various fields, such as circuit design, aerodynamic design, and engineering in general \cite{barthelemy1995automatic, circuits, airfoils}, its application to particle and nuclear physics detectors remains challenging. This difficulty primarily arises from the intrinsic stochasticity introduced by the quantum nature of the physical processes involved. As a result, building differentiable pipelines becomes complex, since most Monte Carlo (MC) particle simulators, such as \textsc{GEANT4} \cite{agostinelli_2003_geant4, allison_2006_geant4_developments, allison_2016_recent_geant4}, are not inherently differentiable. Although efforts have been made to overcome this limitation for specific physical processes \cite{Baydin02012021}, the full optimization of particle detectors remains an open challenge. To date, only a handful of studies have addressed this problem \cite{Baydin02012021, Ratnikov_2020, Cisbani_2020, Koser_IsoDAR, ML_Accelerator_Physics, DORIGO2020100022}.

This approach is becoming feasible due to the ongoing efforts of collaborative research such as MODE Collaboration \cite{Baydin02012021}, which aims to utilize DP and AD for optimizing detector designs in particle physics by developing modular and customizable differentiable pipelines for the optimization of objective functions. The recent success of MODE Collaboration in applying these concepts to fully model a muon tomography system and develop a package called TomOpt \cite{tomopt} led to this study, where this package is adapted to a gaseous detector designed for heavy-ion tracking and imaging: the Optical Parallel-Plate Avalanche Counter (O-PPAC).

The basic design of the O-PPAC, introduced by Cortesi et al.~in Reference \cite{cortesi18}, consists of two parallel squared electrodes separated by a small 3 mm gap filled with a low-pressure scintillating gas, with an array of small, collimated photo-sensors along the edges of the gap. The position of an impinging particle is reconstructed using the information provided by the distributions of detected photons along the edges of the gas gap. 

The main goal of this study is to identify the optimal values of two detector parameters: the pressure of the scintillating gas ($p$) and the length of the collimator ($L$), both of which affect the spatial resolution. To achieve this, a differentiable pipeline is developed to minimize the reconstruction error as a function of the detector parameters. This is performed by developing a surrogate model that replicates an existing \textsc{GEANT4} simulation of this detector fully described in Reference \cite{cortesi18}, where 5.5 MeV alpha particles traverse the detector perpendicularly to the electrodes.

The surrogate acts as a differentiable approximation of the detector's behavior, enabling a gradient-based optimization using AD. In this study, PyTorch's~\cite{PyTorch} AD features are employed.

The structure of the document is as follows: Section \ref{sec:materials_and_methods} briefly introduces the operational principle of the detector and provides a detailed explanation of the optimization methods applied. Section \ref{sec:results} presents the results obtained from the optimization process. Lastly, Section \ref{sec:discussion} discusses the conclusions drawn from this work.

\section{Materials and Methods}
\label{sec:materials_and_methods}

\subsection{The O-PPAC Detector}

Despite being an older detector concept, position-sensitive parallel-plate avalanche counters (PPACs) remain widely used today in various subatomic physics applications, particularly for heavy-ion position and timing measurements \cite{PhysRevAccelBeams.27.044801}.  

As discussed in Reference \cite{cortesi18}, unlike conventional PPACs that use charge division or delay-line methods, electroluminescence-based detectors benefit from high-sensitivity solid-state photosensors, such as SiPMs, which provide better signal-to-noise ratios and energy resolution, making them well suited for heavy-ion tracking and imaging.

One example of this approach is the O-PPAC design \cite{cortesi18}, which consists of two parallel electrodes separated by a narrow 3 mm gap. This gap is filled with a low-pressure scintillating gas, such as CF4, known for its high electroluminescence light yield. Along the edges of the avalanche gap, arrays of small, collimated SiPMs are strategically arranged to maximize light collection and improve detection efficiency.

\begin{figure}[h] 
\centering
\includegraphics[scale=0.5]{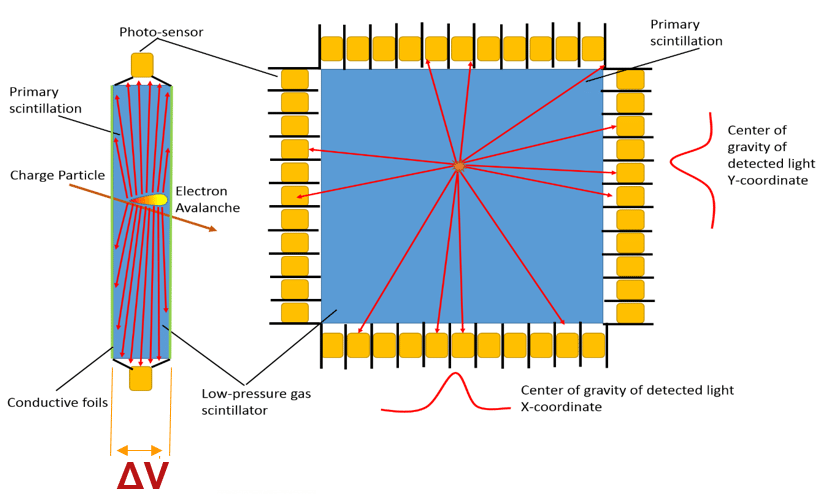} 
\caption{{Schematic} 
 representation of the operational principle of the Optical Parallel-Plate Avalanche Counter (O-PPAC). The photosensors are highlighted in yellow, the collimator in black, and the gas volume in blue. The left side shows a lateral view of the O-PPAC, illustrating the avalanche process initiated when a particle traverses the active medium, with the scintillating photons and their trajectories highlighted in red. On the right, the front view of the detector is presented, demonstrating the effect of collimation. Image adapted from Reference \cite{cortesi18}.} \label{fig/operational_OPPAC} 
\end{figure}

When an ionizing particle crosses the active volume, it releases a small amount of energy in the form of ionization electrons, which are multiplied in the gas by a uniform electric field established between the two metalized parallel plates. The scintillation light emitted during the avalanche process, known as secondary scintillation or electroluminescence,  is reflected back and forth by the two metallized electrode foils and guided to be recorded by the arrays of collimated photo-sensors (see Figure~\ref{fig/operational_OPPAC}). The collimation is crucial for the precise localization of the impinging particles, as it narrows down the detected photon distributions, so its peak is more heavily weighted near the position of the avalanche. 

In this study, the parameters of interest were the pressure of the scintillating gas and the length of the collimator, as they are crucial parameters for the characteristics of the detected photon distributions.

The pressure of the gas is directly related to the electroluminescence yield, as explained in Reference~\cite{cortesi2016}. As a general rule, higher pressure implies a higher electroluminescence yield, which translates into a higher number of detected photons in each event and, therefore, higher statistics in the detected photon distributions.

As the width of the collimator walls (1 mm) is determined by the manufacturing constraints of the SiPMs selected by the detector designers~\cite{cortesi18} and alternative technologies were not explored in this study, the optimization efforts were focused on the other relevant parameter regarding collimation: the collimator length. The optimal collimator length results from a trade-off between achieving more accurate light spot localization with larger $L$ and obtaining higher statistics with smaller $L$.

It is also worth mentioning that the number of SiPMs was fixed to 33 per wall, the SiPM effective area was $2\times 3$ mm$^2$, the pitch was 3 mm long, and the cathode and anode metalized foils (Al) were characterized by a reflectivity of 90$\%$. The remaining simulation parameters not mentioned in this study are exactly as detailed in Reference~\cite{cortesi18}, with the exception of the collimator length and pressure, which are studied in this work.

\subsection{End-to-End Optimization}

As mentioned earlier, applying gradient-based optimization techniques requires the detector response to be in a differentiable form. However, \textsc{GEANT4} simulations remain non-differentiable, which makes it challenging to directly apply gradient-based methods.

To overcome this, approximate surrogate models can be used as an alternative to Monte Carlo simulations \cite{NEURIPS2020_a878dbeb}. These models, which are typically obtained by employing some form of supervised training on events previously generated by the MC simulator, not only enable differentiation but also offer practical benefits: once trained, they run significantly faster than a gradient-aware MC simulation and often provide a smoother approximation, which is better suited for gradient-based optimization~\cite{Aehle:2882304,Kasim_2022}.

In this study, a surrogate model is obtained by training a Neural Network (NN) on a grid of MC simulated points of the parameter space. After the training step, a differentiable model that can be inserted into a differentiable pipeline just as a closed-form expression is obtained, which is the key to this approach. Specifically, once trained, this NN will predict the reconstructed position of the particle as a function of the detector parameters and its initial position. 

Once the model is trained, the next step is to build an optimization loop that will minimize an objective function by iteratively updating the values of the detector parameters according to the gradients of this function at each step.

In the following subsections, each step involved in the optimization of the detector is described in detail.

\subsubsection{Simulation}

The first step to optimize the detector is to generate a set of datasets corresponding to a grid of the interest parameters and the beam position. As stated before, our goal is to simulate a reduced number of datasets and then train an NN model that extrapolates all the possible configurations with a differentiable model. For this, the \textsc{GEANT4} simulation described in Reference~\cite{cortesi18} is employed.

The grid of parameters employed for the simulation is provided in Table \ref{tab:simulation}. It is important to highlight that each dataset consisted of 10,000 events, each involving a 5.5 MeV alpha particle entering the detector perpendicular to the parallel plates.

\begin{table}[H] 
\caption{{Summary} 
 of the simulated parameter values. The simulation explored all possible combinations of the listed parameters, resulting in a total of 2025 unique configurations.\label{tab:simulation}}
\newcolumntype{C}{>{\centering\arraybackslash}X}
\begin{tabularx}{\textwidth}{CC}
\toprule
\textbf{Parameter} & \textbf{Simulated Values} \\
\midrule
Pressure (Torr) & 10,  20,  30,  40,  50  \\
Collimator length (mm) & {5}
,  16.25,  27.5,  38.75,  50.  \\
X position of the beam (cm) & {$-$}
4, $-$3, $-$2, $-$1, 0, 1, 2, 3, 4 \\
Y position of the beam (cm) & $-$4, $-$3, $-$2, $-$1, 0, 1, 2, 3, 4 \\
\bottomrule
\end{tabularx}
\end{table}

As can be observed, the pressure and the collimator length were bound to lie within a certain interval. The minimum pressure value is justified by the photon distribution statistics, as trial simulations showed that below 10 Torr, few to no photons were detected.  The upper limit of 50 Torr is determined by the fact that higher pressure requires an increase in the voltage between the parallel plates, which cannot exceed a certain maximum.

In a similar manner, the lower limit of the collimator length is set taking into account the poor precision obtained, as the dispersion of the distribution when  $L<5$ mm is too high. The upper limit is set due to the poor statistics obtained when $L>50$ mm.

\subsubsection{Reconstruction of the Position}

\begin{figure}[h]
\centering
    \includegraphics[width=0.78\linewidth]{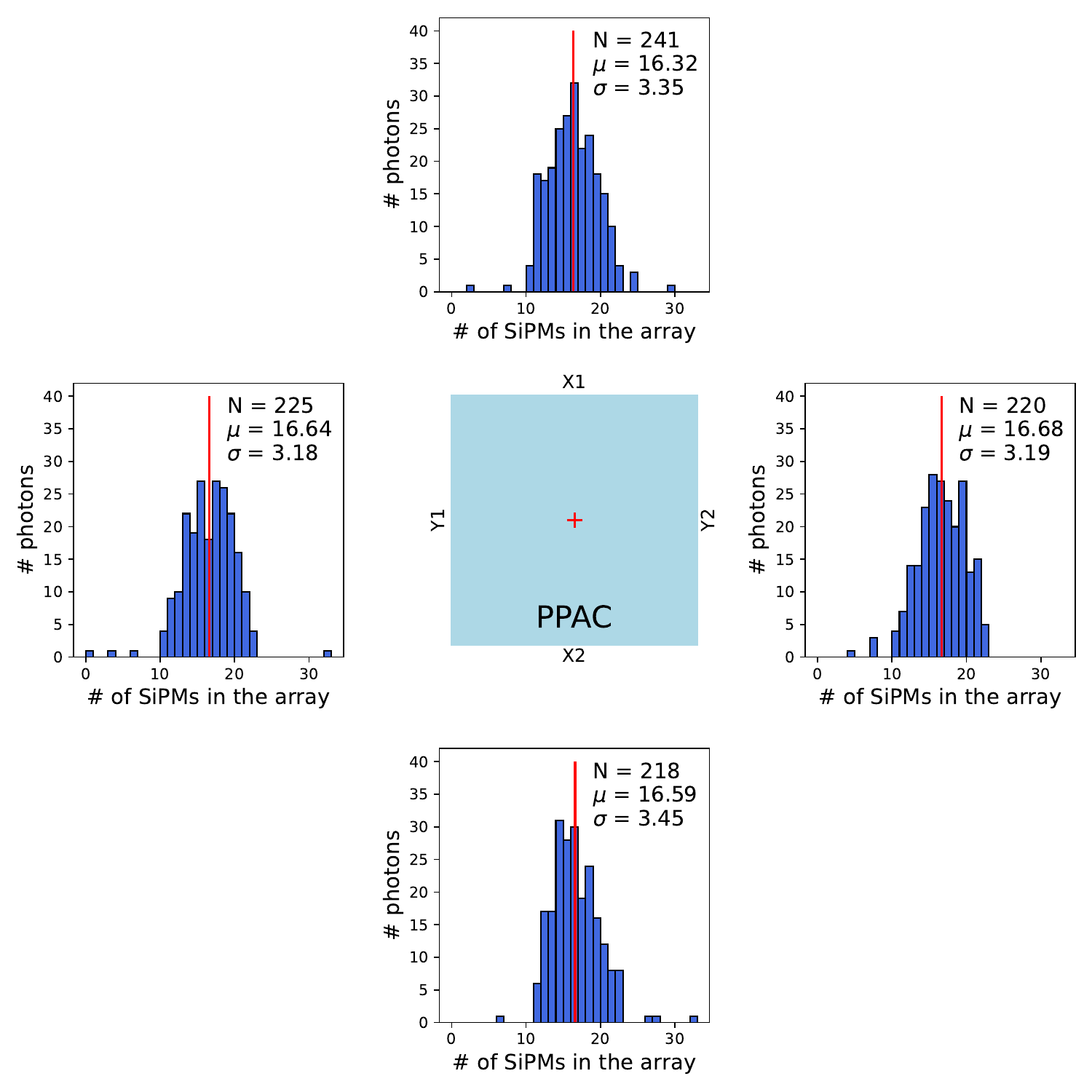}
    \caption{Illustration of the reconstruction process for a simulated event induced by a 5.5 MeV alpha particle at the center of the parallel plates. The photon counts registered by each photosensor are used to generate four distributions, one for each wall. The resulting plots provide the statistical information required for reconstruction, with the mean of each distribution highlighted in red. The particle's position along each axis is determined by analyzing the signals from opposing walls. }
    \label{fig:reconstruction}
\end{figure}

The reconstruction of the avalanche location, which corresponds to the position of the charged-particle crossing the detector volume ($\hat{x}$, $\hat{y}$), is achieved by combining the data recorded by the four photo-sensor arrays located in each wall of the PPAC, as illustrated in Figure \ref{fig:reconstruction}.

This task can be performed in several ways, but the simplest is to compute the arithmetic mean between the photon distribution peaks recorded  by each pair of opposing arrays weighted by the total number of detected photons and the dispersion of each distribution~\cite{cortesi18}:

\begin{equation}
\hat{x}= \left(\frac{\mathrm{P}_{\mathrm{x} 1} \cdot \mathrm{N}_{\mathrm{x} 1}}{\sigma_{\mathrm{x} 1}}+\frac{\mathrm{P}_{\mathrm{x} 2} \cdot \mathrm{N}_{\mathrm{x} 2}}{\sigma_{\mathrm{x} 2}}\right) {\Big/}\left(\frac{\mathrm{N}_{\mathrm{x} 1}}{\sigma_{\mathrm{x} 1}}+\frac{\mathrm{N}_{\mathrm{x} 2}}{\sigma_{\mathrm{x} 2}}\right)
\label{eq:reco_X}
\end{equation}

\begin{equation}
\hat{y}=\left(\frac{\mathrm{P}_{\mathrm{y} 1} \cdot \mathrm{N}_{\mathrm{y} 1}}{\sigma_{\mathrm{y} 1}}+\frac{\mathrm{P}_{\mathrm{y} 2} \cdot \mathrm{N}_{\mathrm{y} 2}}{\sigma_{\mathrm{y} 2}}\right) {\Big/}\left(\frac{\mathrm{N}_{\mathrm{y} 1}}{\sigma_{\mathrm{y} 1}}+\frac{\mathrm{N}_{\mathrm{y} 2}}{\sigma_{\mathrm{y} 2}}\right),
\label{eq:reco_Y}
\end{equation}
where P and $\sigma$ correspond to the mean and the standard deviation of the distribution on each wall and N is the total number of photons detected in each wall.

\subsubsection{Surrogate Model}

As stated before, the algorithm employed to parametrize the reconstruction is a Neural Network developed employing PyTorch \cite{PyTorch}. This particular NN receives four inputs ($x$, $y$, $p$, $L$) and predicts two outputs ($\hat{x}$, $\hat{y}$), i.e., the reconstructed position. 

In order to optimize, train, and evaluate the model, the simulated events were divided into different datasets, as illustrated in Table \ref{tab:datasets}. For instance, a small fraction (around {$5 \times 10^4$} 
 events) was employed for hyperparameter tuning, a bigger dataset for training (around 1 million), and the remaining MC events were used for the evaluation of the NN.

\begin{table}[H] 
\caption{{Number} 
 of simulated events used in the different phases of the surrogate model development process, including hyperparameter tuning, training, and evaluation.}
\label{table:data_counts}
\newcolumntype{C}{>{\centering\arraybackslash}X}
\begin{tabularx}{\textwidth}{CCCC}
\toprule
\textbf{Dataset} & \textbf{Training} & \textbf{Validation} & \textbf{Evaluation} \\ \midrule
Hyperparameter tuning & $2.5 \times 10^4 $ & $2.5 \times 10^4 $ & --- \\
Training & $5 \times 10^5 $ & $5 \times 10^5 $ & --- \\
Evaluation & --- & --- & $2 \times 10^7$ \\ \bottomrule
\end{tabularx}

\label{tab:datasets}
\end{table}

It is worth mentioning that the splitting in the different datasets was made maintaining the proportion of the different parameters, so that all combinations are equally represented in each dataset.

To optimize the performance of the NN, a hyperparameter tuning process was conducted to identify the optimal set of model parameters. The package employed for hyperparameter tuning was Optuna~\cite{optuna}, which utilizes the Tree-structured Parzen Estimator (TPE) as its default algorithm for sampling candidates in the search space. TPE is a widely utilized Bayesian optimization method that iteratively constructs two probability density functions: one for the hyperparameters of successful trials (i.e., favorable configurations), and another for unsuccessful trials (i.e., unfavorable configurations). The algorithm employs these probability density functions to sample new hyperparameters that are likely to improve the objective function. This method has demonstrated greater efficiency and effectiveness compared to other well-established hyperparameter tuning methods, such as Grid Search or Randomized Search.

The hyperparameter tuning process was conducted in two distinct phases: the initial phase focused on architecture optimization, while the subsequent phase aimed at further enhancing performance and regularization.

To achieve this objective, the hidden size, number of hidden layers, and learning rate are optimized in the initial phase. The considered values of these hyperparameters, along with the optimal parameters obtained from 100 trials, are detailed in Table~\ref{tab:tuning_results}.

\begin{table}[h] 
\caption{Results from the first hyperparameter tuning with Optuna, showing the best trial out of 100. The table lists the hyperparameters considered, their tested values, and the optimal values based on model performance.}
\label{tab:tuning_results}
\newcolumntype{C}{>{\centering\arraybackslash}X}
\begin{tabularx}{\textwidth}{CCC}
\toprule
\textbf{Hyperparameter} & \textbf{Considered Values} & \textbf{Best Trial} \\ \midrule
Hidden size             & 32, 64, 128, 256, 512   & 64                  \\ 
Number of layers        & 2, 3, 4, 5, 6, 7, 8   & 3                   \\ 
Learning rate           & From 0.001 to 0.1                & 0.0352  \\ 
\bottomrule
\end{tabularx}
\end{table}
   
As stated before, the second phase of the hyperparameter tuning was focused on further improving the performance of the NN, on regularization and stability. For this purpose, hyperparameters like dropout, optimizer, learning rate scheduler, batch normalization, and activation function were studied. Again, the considered options and the Optuna's best trial out of 1000 trials is specified in Table~\ref{tab:tuning_results_round2}.

\begin{table}[h] 
\caption{Results from the second hyperparameter tuning with Optuna, showing the best trial out of 1000. The table lists the hyperparameters considered, their tested values, and the optimal values based on model performance.}
\label{tab:tuning_results_round2}
\newcolumntype{C}{>{\centering\arraybackslash}X}
\begin{tabularx}{\textwidth}{ccC}
\toprule
\textbf{Hyperparameter} & \textbf{Considered Values}& \textbf{Best Trial}\\ \midrule
Dropout & 0, 0.1, 0.2, 0.3, 0.4, 0.5 & 0 \\ 
Optimizer                      & Adam, NAdam, Adamax                            & Adamax                             \\ 
Step Size Scheduler            & 1, 2, 3, 4, 5                                  & 2                                  \\ 
Gamma Scheduler                & 0.9, 0.99, 0.999                               & 0.9                                \\ 
Use Batch Normalization        & True, False                                   & False                              \\ 
Activation Function            & ReLU, SELU, ELU, LeakyReLU                  & ELU                               \\ \bottomrule
\end{tabularx}
\end{table}

In Table~\ref{tab:tuning_comparative}, the evaluation of the models for each round of hyperparameter tuning is shown. Specifically, the root mean squared error (RMSE) between the reconstructed position and the NN prediction is evaluated for the large dataset. As shown, the evaluation yields an RMSE of 0.035 cm  both for the x- and y-axes after the second phase, which is an improvement with respect to the model resulting from the first phase of hyperparameter~tuning.

\begin{equation}
    \text{RMSE} = \sqrt{\frac{1}{n} \sum_{i=1}^{n} |\vec{x}_{NN}^i - \hat{\vec{x}}^i|^2}
\end{equation}

\begin{table}[H] 
\caption{NN prediction error (RMSE) after each step of the hyperparameter tuning process. The table shows the root mean square error (RMSE) for both the x- and y-coordinates at each step of the tuning.}
\label{tab:tuning_comparative}
\newcolumntype{C}{>{\centering\arraybackslash}X}
\begin{tabularx}{\textwidth}{CCC}
\toprule
\textbf{Step} & \textbf{RMSE(\boldmath{$\hat{x}, x_{NN}$}) [cm]} & \textbf{RMSE(\boldmath{$\hat{y}, y_{NN}$} 
) [cm]} \\ \midrule
1 & 0.040 & 0.041\\
2 & 0.035  & 0.035\\
 \bottomrule
\end{tabularx}
\end{table}

\begin{figure}[H]
\centering
    \includegraphics[width=0.65\linewidth]{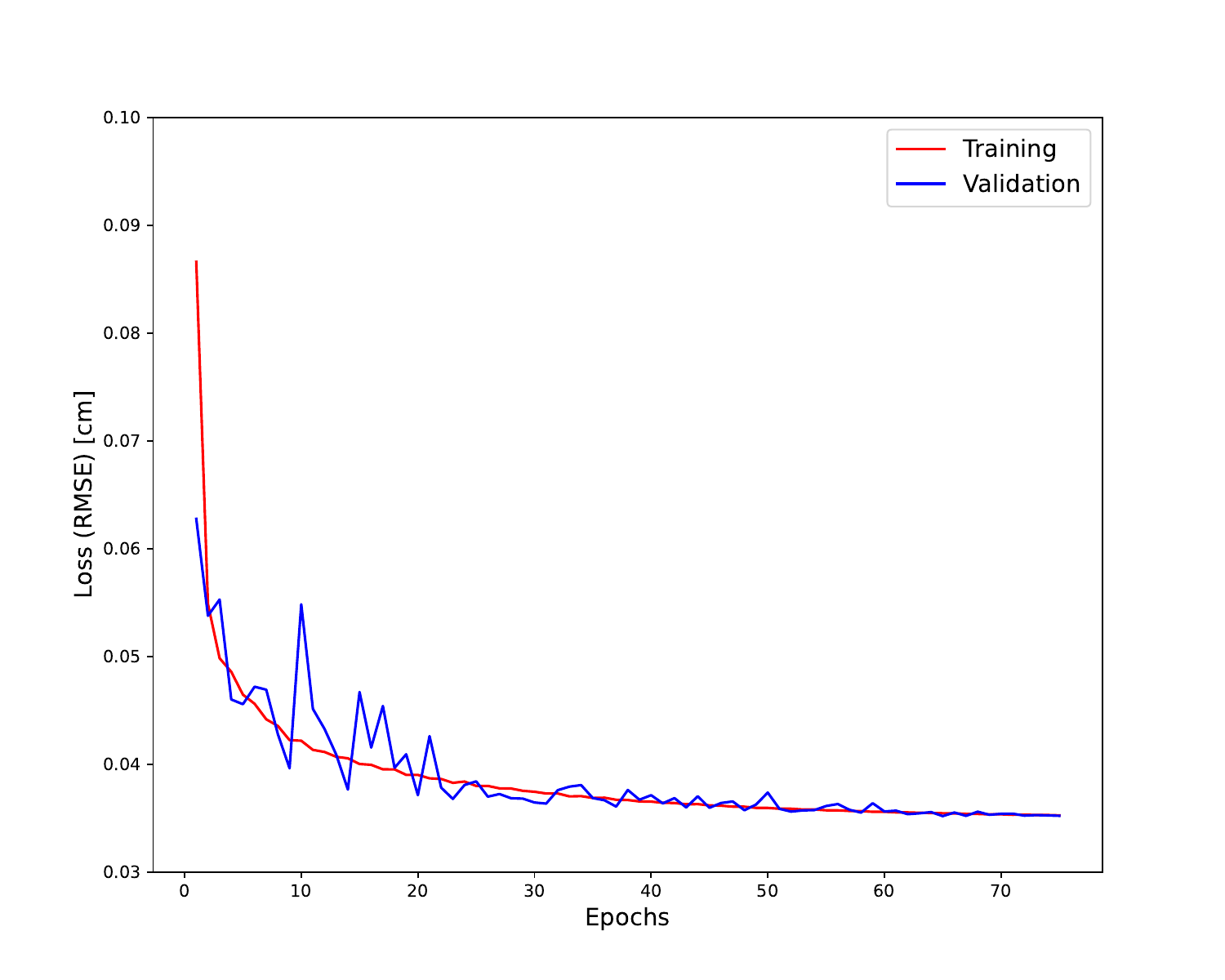}
    \caption{{Learning} 
 curve of the NN surrogate model following the second phase of hyperparameter tuning. The plot illustrates the model's training progress, with early stopping applied using a tolerance ($\delta$) of 0.001 and a patience of 10 epochs to prevent overfitting and reduce computational~time.}
    \label{loss}
\end{figure}

Additionally, the learning curve for the final model is illustrated in Figure \ref{loss}. As can be seen, the validation loss curve exhibits an erratic behavior during the first 20 epochs, but it stabilizes afterwards, with the training loss value closely matching the validation loss, which suggests a low level of overfitting.

\subsubsection{Optimization}

This section provides an overview of the optimization process and its implementation in the repository~\cite{repo}. It describes the core components of the optimization package and offers a detailed explanation of the optimization loop. It is important to note that this code is derived from TomOpt package~\cite{strong_2024_10673885}.

Below are the key components necessary for implementing the optimization, along with their descriptions:

\begin{enumerate}[label=,leftmargin=0em,labelsep=4mm]
\item \textbf{{Alpha} 
 Batch:}
\end{enumerate}

In order to perform the optimization loop, a class named AlphaBatch is built. This class has the main goal of producing batches of alpha beams in the form of random Pytorch tensors. The random numbers are generated uniformly  both in the x- and y-axes in the interval $[-4, 4]$ cm, as the surrogate model is trained on that interval.

\begin{enumerate}[label=,leftmargin=0em,labelsep=4mm]
\item {\bf{{Volume}:}} 
\end{enumerate}

The Volume refers to the detector volume itself, represented as a class defined by the detector parameters at each step in the optimization loop. It is initialized with values for pressure and collimator length, which are the parameters of interest.

The primary purpose of this class is to accept a specific detector configuration and an alpha batch, then to predict the reconstructed position for each impinging alpha particle using the surrogate model.

Moreover, the Volume class includes a method that ensures the parameter values remain within a predefined interval during the optimization loop. If a parameter exceeds the boundaries of this interval during training, its value is clamped to the nearest limit. This functionality is crucial for the optimization process, as the surrogate model is trained only within specific parameter ranges and should not be extrapolated outside these intervals.

\begin{enumerate}[label=,leftmargin=0em,labelsep=4mm]
\item {\bf{{Objective} function:}} 
\end{enumerate}

The objective function is the metric minimized during the optimization process. It can be designed to prioritize parameter combinations that enhance both the detector's performance and cost-efficiency. In this study, the objective function was performance-focused, using the root mean square error (RMSE) between the reconstructed position and the beam position as the optimization criterion.

\begin{enumerate}[label=,leftmargin=0em,labelsep=4mm]
\item{\bf{{Volume} Wrapper and optimization Loop:}} 
\end{enumerate}

\begin{figure}[h]
\centering
    \includegraphics[width=0.7\linewidth]{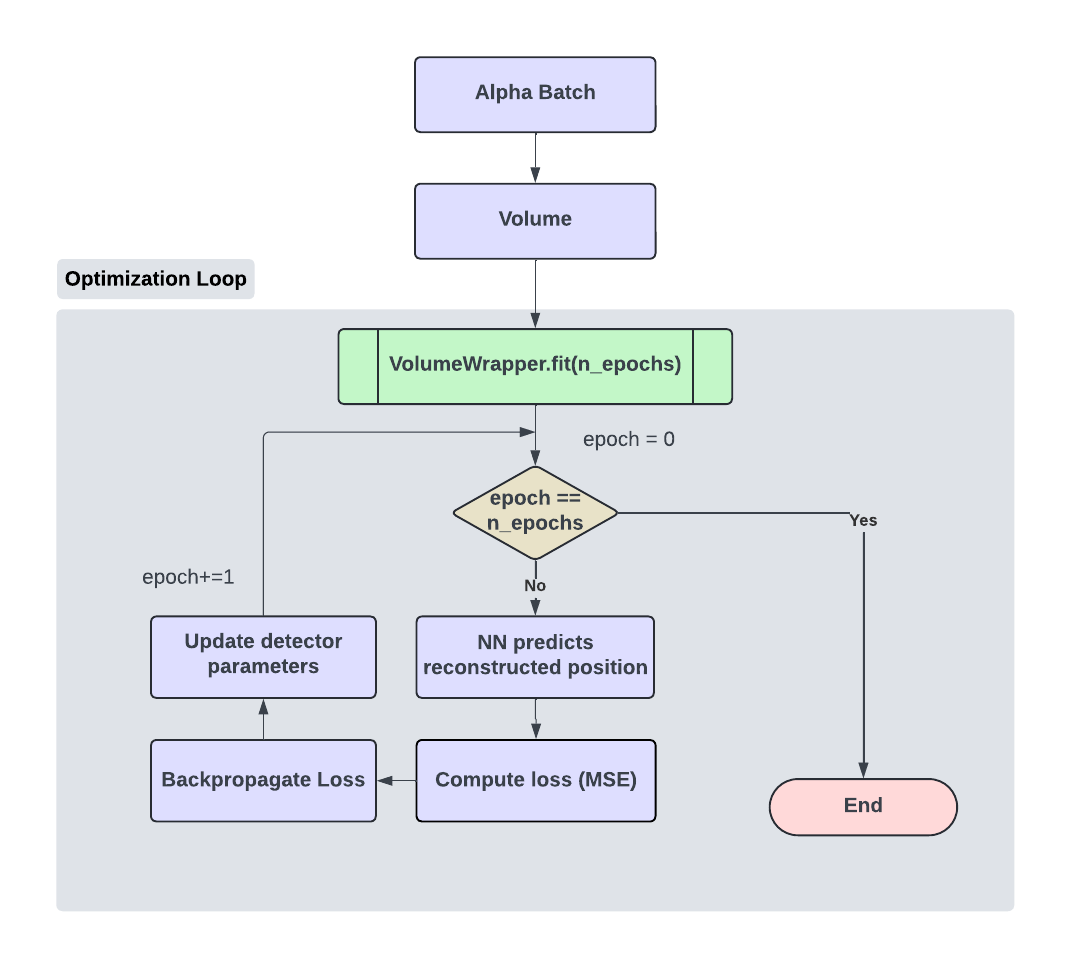}
    \caption{Breakdown of the detector optimization loop, outlining the process of initializing and updating the parameters ($p$
 and $L$) based on the gradient of the loss function. The loop runs through multiple epochs, where each iteration includes model prediction, loss calculation, and parameter updates to minimize the loss function.}
    \label{fig/optloop}
\end{figure}

The Volume class includes a Wrapper that contains the fit method, responsible for optimizing the detector by determining the optimal configuration based on the NN reconstruction model. This optimization is conducted within a loop designed to iteratively identify the best detector configuration. The structure of this loop is illustrated in Figure~\ref{fig/optloop}.

As explained in Figure~\ref{fig/optloop}, the first step is to initialize an alpha batch with N alphas generated in random positions in the interval $[-4, 4]$ cm both in the x- and y-axes. Then, the volume wrapper class is initialized with the alpha batch and initial  values of the pressure ($p$) and the collimator length ($L$).

Before initializing the loop, the optimizer is initialized to update the values of both parameters based on the gradient of the loss function, according to a previously specified learning rate. In this study, the Adam optimizer was employed with a learning rate of 0.1 and default values for ($\beta_1$, $\beta_2$).

At each epoch, the NN reconstruction model receives the inputs, which are the alpha batch and the initial detector configuration. The model then predicts the reconstructed position of each alpha particle.

These predicted positions, along with the real beam positions, are fed into the loss function. The next step involves backpropagating the loss to compute the gradients of this function with respect to the parameters to be optimized ($p, L$). Once the gradients are obtained, the optimizer takes a step, updating the detector parameters according to the computed gradients.

This process is repeated over a certain number of epochs until all values stabilize and, as a result, the detector configuration that minimizes the loss function is obtained.

\section{Results}
\label{sec:results}

\subsection{Optimization for a Random Initial Configuration}
\label{sec:random_config}

\begin{figure}[h]
\centering
\subfloat[\centering]{\includegraphics[width=6.5cm]{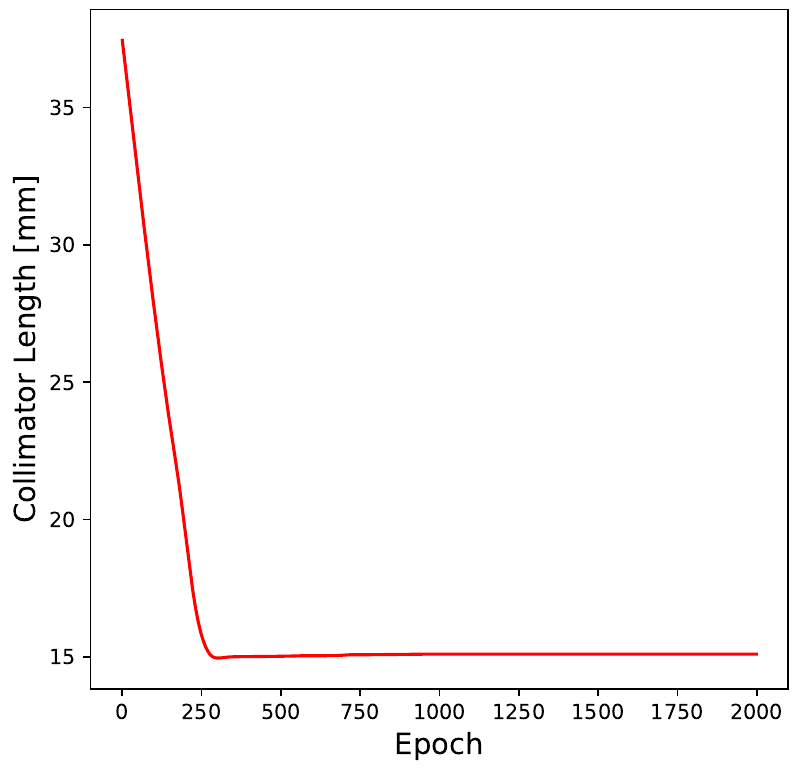}}
\hspace{0.5cm}
\subfloat[\centering]{\includegraphics[width=6.5cm]{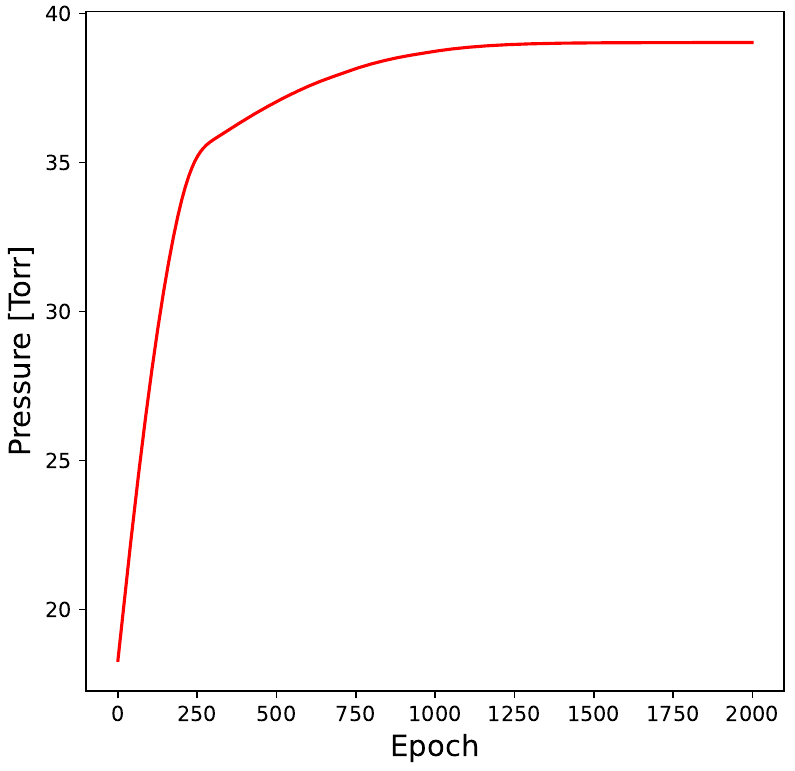}}\\
\vspace{0.5cm}
\subfloat[\centering]{\includegraphics[width=6.5cm]{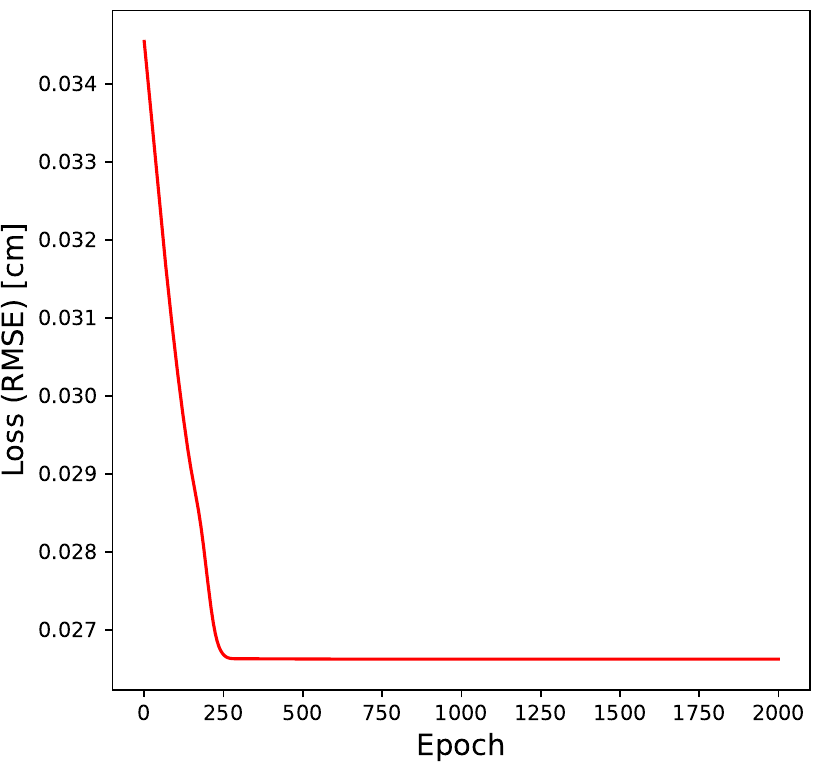}}
\caption{Evolution of the (\textbf{a}) collimator length, (\textbf{b}) pressure, and (\textbf{c}) loss function throughout the optimization loop for a random initial configuration of pressure and collimator length. The figures show the progression of these parameters as the optimization loop iterates, with the values adjusting to minimize the loss function and improve the detector configuration.\label{fig:single_opt}}
\end{figure}

First, a single optimization loop was carried out with an alpha batch of 10,000 alphas located at random positions and a random initial configuration of the detector during 1500 epochs. In Figure~\ref{fig:single_opt},  the evolution of the loss function, as well as the evolution of the parameters throughout the optimization loop is presented. The optimal values of the pressure and the collimator length found with this method are presented in Table~\ref{tab:results_random_config}.

\begin{table}[h] 
\caption{Optimal values of pressure ($p$) and collimator length ($L$) found for a random initial configuration and an Alpha Batch of $10^5$ particles, randomly distributed following a uniform distribution in the interval  $[-4, 4]$ cm.\label{tab:results_random_config}}

\newcolumntype{C}{>{\centering\arraybackslash}X}
\begin{tabularx}{\textwidth}{CCC}
\toprule
\textbf{Parameter} & \textbf{Found Optimal Value} \\
\midrule
Pressure          &  39.03 Torr \\
Collimator length &  15.11 mm  \\
\bottomrule
\end{tabularx}
\end{table}

\begin{figure}[h]
\centering
\includegraphics[width=7.5cm]{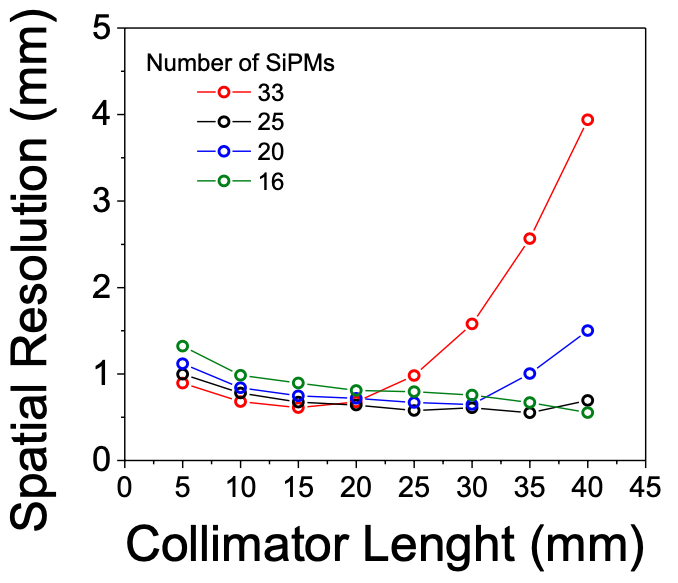}
\caption{Position resolution as function of the collimator length, for SiPMs arrays of different granularity---the number of photo-sensor per array ranges from 12 to 33 elements. The pressure was fixed to 30 Torr. Image from Reference~\cite{cortesi18}.}
\label{fig:dcol_cortesi18}
\end{figure}

The result for the collimator length agrees with a previous result obtained in \mbox{Reference~\cite{cortesi18}}, where traditional optimization techniques were employed, involving simulations of the detector under different parameter combinations, assessing their performance, and selecting the optimal one. In Figure~\ref{fig:dcol_cortesi18}, the spatial resolution as a function of the collimator length for different number of SiPMs per wall is illustrated. The minimum of the curve for 33 SiPMs, which is the value used in this study, corresponds to a collimator length of approximately 15 mm. This is very close to the value obtained through automatic optimization for a similar pressure (30 Torr).

\newpage

\subsection{ Optimization for a Detector Configuration Grid}

To verify whether the previous result is independent of the initial configuration of the parameters---which would indicate that the identified minimum is the absolute minimum of the objective function---the optimization loop was repeated across a grid of initial detector configurations. The values considered for both parameters are detailed in Table~\ref{tab:detector_configurations}. This grid translates into 400 different configurations and, therefore, 400 optimization loops that are carried out just as described in Section \ref{sec:random_config}. 

\begin{figure}[H]
\centering
\subfloat[\centering]{\includegraphics[width=6.8cm]{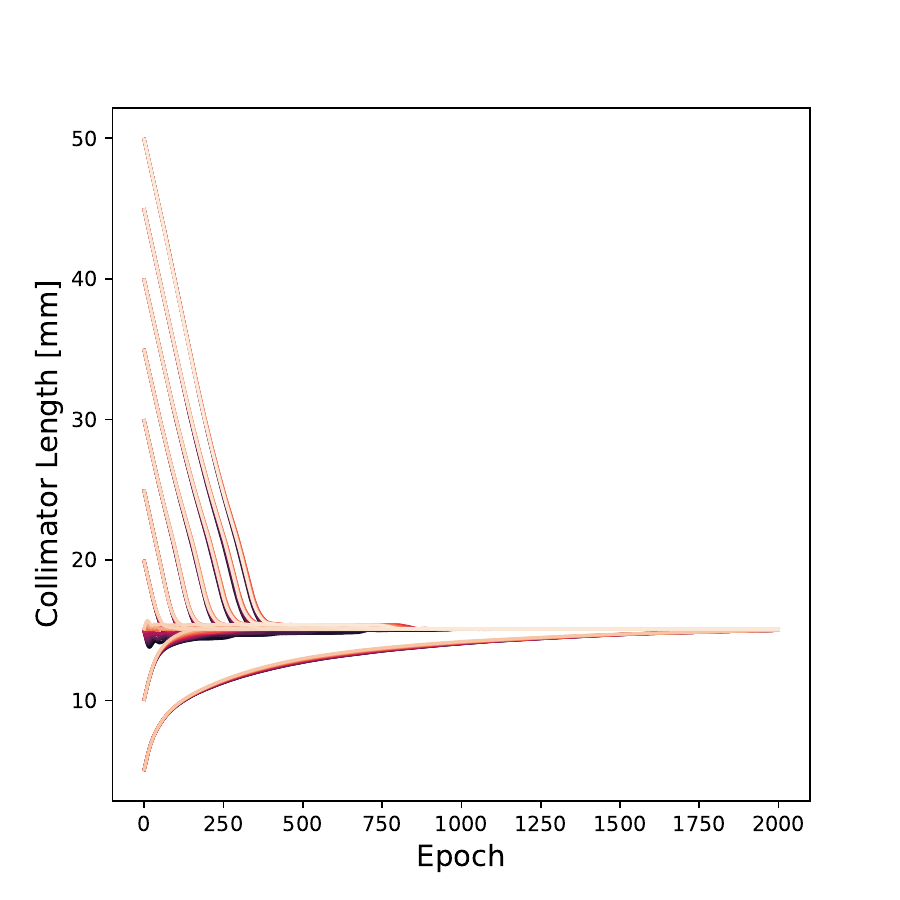}}
\hspace{0.5cm}
\subfloat[\centering]{\includegraphics[width=6.8cm]{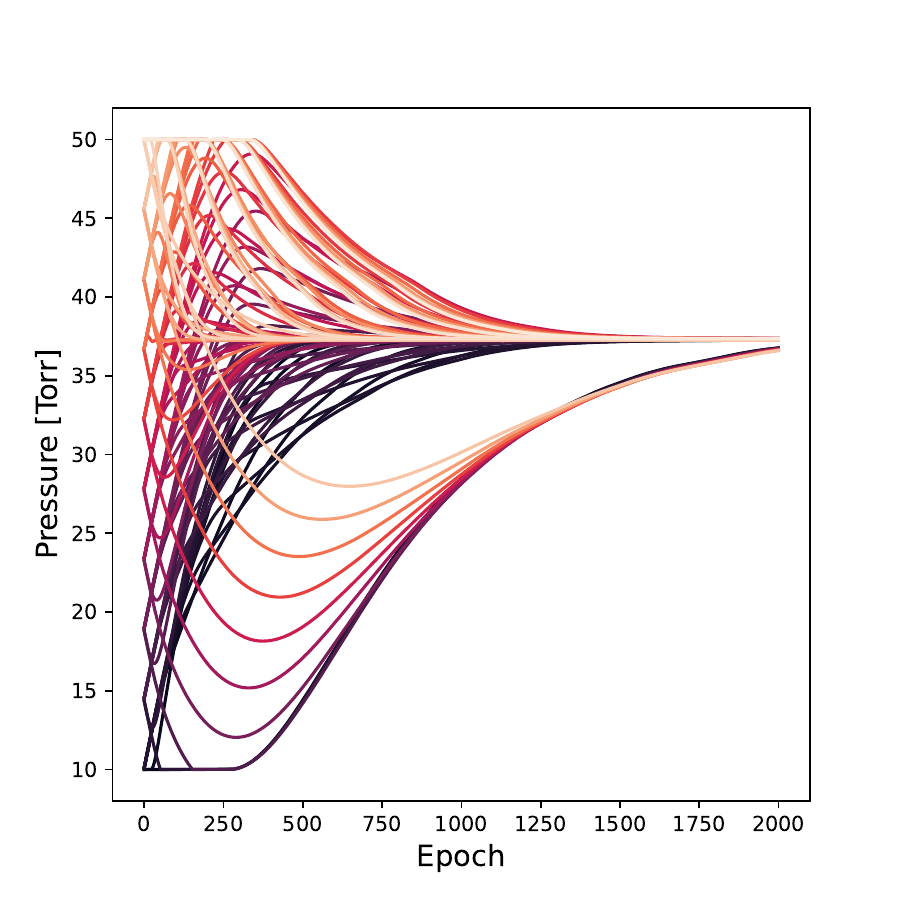}}\\
\vspace{0.5cm}
\subfloat[\centering]{\includegraphics[width=6.8cm]{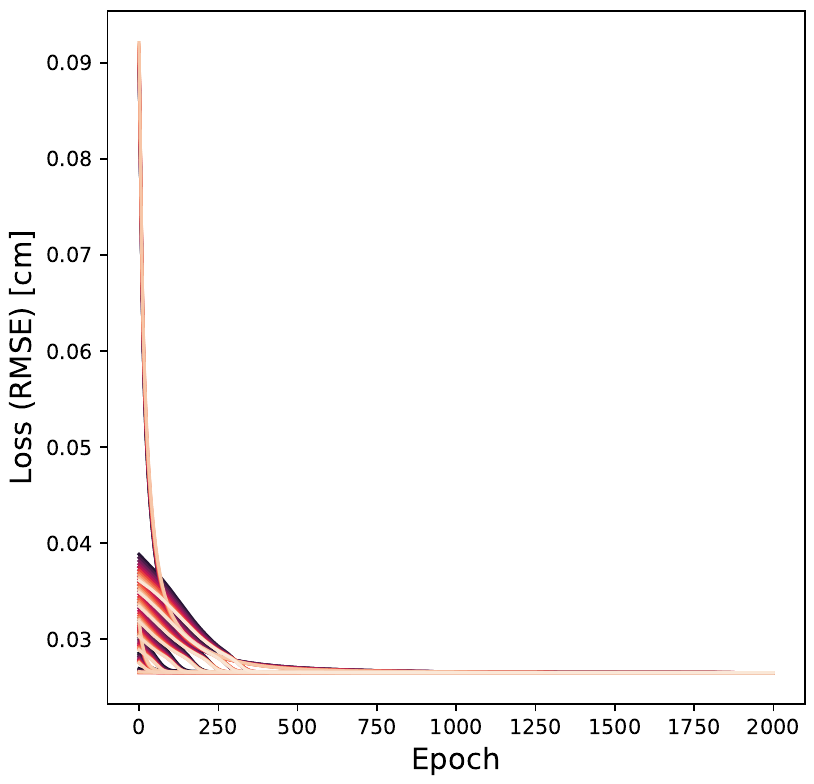}}
\caption{Evolution of the (\textbf{a}) collimator length, (\textbf{b}) pressure, and (\textbf{c}) loss function throughout the optimization loop for a grid of initial configurations of pressure and collimator length. Despite different starting points, the optimization process consistently converges to the same final result, suggesting that the found minimum of the loss function is the absolute minimum.\label{fig:grid_opt}}
\end{figure}

As shown in Figure~\ref{fig:grid_opt}, all configurations converge to the same optimal values for pressure and collimator length, indicating that this minimum of the loss function is the absolute minimum within the studied range.

\begin{table}[h] 
\caption{Initial values of the parameter grid considered for the optimization loop, specifying the range of values tested for both the pressure ($p$) and collimator length ($L$).\label{tab:detector_configurations}}
\newcolumntype{C}{>{\centering\arraybackslash}X}
\begin{tabularx}{\textwidth}{Cc}
\toprule
\textbf{Parameter}       & \textbf{Values}                                                   \\ \midrule
Pressure                 & 20 values uniformly distributed between 10 and 50 Torr            \\
Collimator length        & 20 values uniformly distributed between 5 and 50 mm               \\
\bottomrule
\end{tabularx}
\end{table}

\begin{figure}[H]
\centering
\subfloat[\centering]{\includegraphics[width=9.5cm]{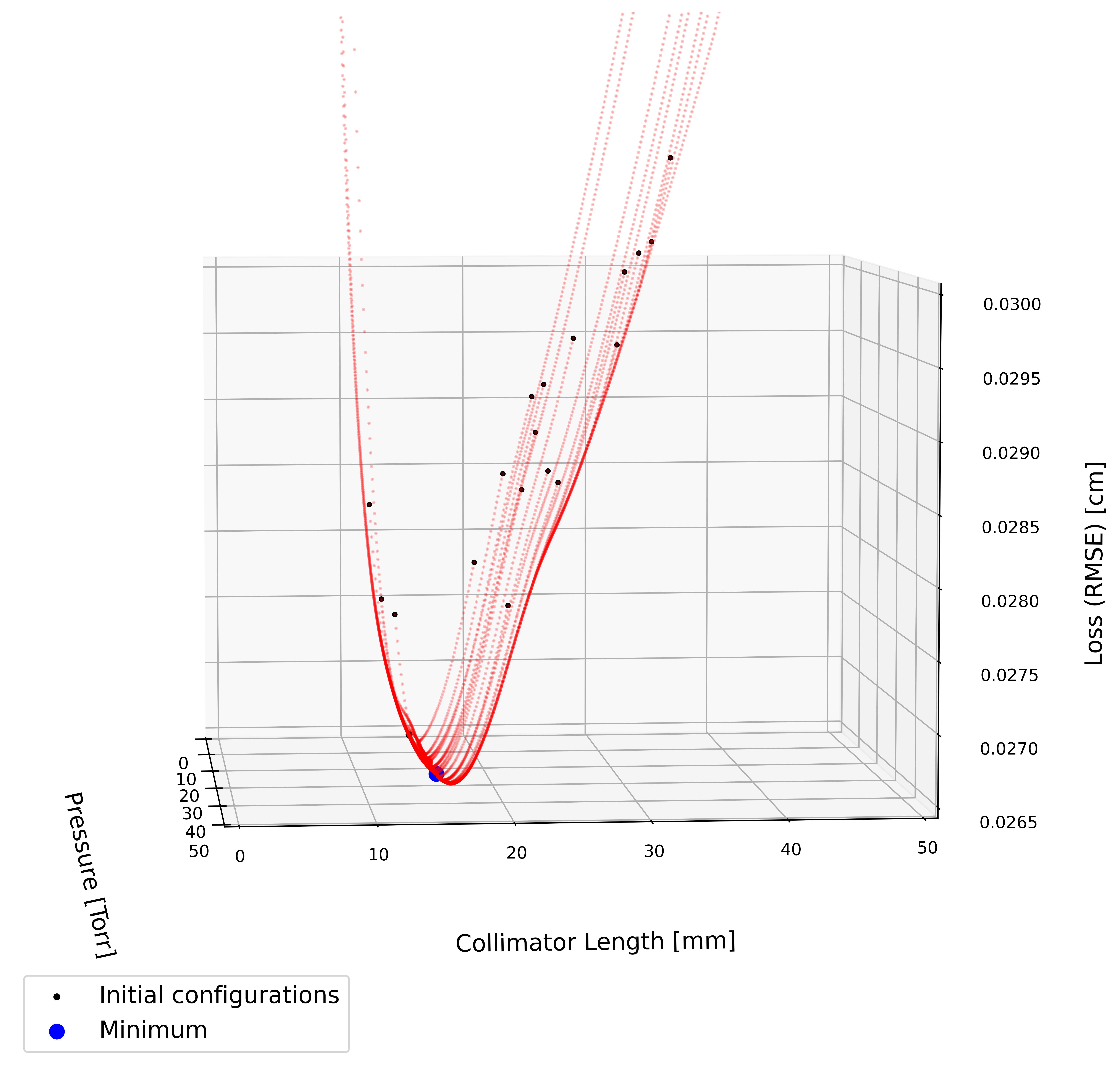}}\\
\subfloat[\centering]{\includegraphics[width=9.5cm]{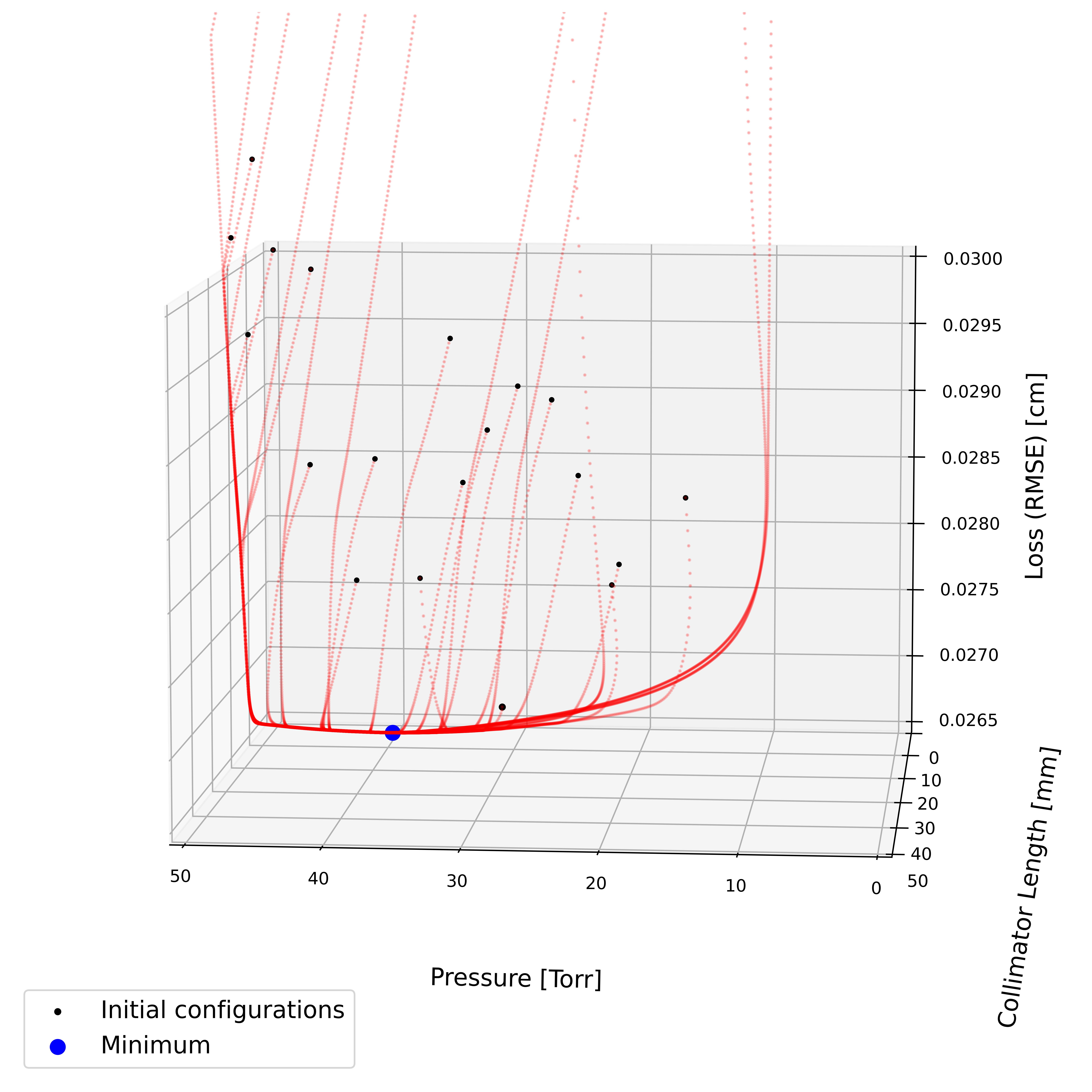}}
\caption{3D representation of several optimization curves along (\textbf{a}) the collimator length axis and (\textbf{b})~the pressure axis, starting from 30 random initial configurations. The minimum of the curve along both axes is clearly observed. Additionally, a significantly higher gradient is observed along the collimator length axis compared to the pressure axis.}
\label{fig:3D_visualization}
\end{figure}

Lastly, Figure~\ref{fig:3D_visualization} presents a 3D representation of several optimization curves. This visualization effectively 'samples' the function $E(p, L)$, where $E$ represents the reconstruction error in the z-axis, $p$ is the pressure in Torr in the x-axis, and $L$ is the collimator length in mm in the y-axis. From this, it can be inferred that the collimator length has a greater impact on the reconstruction error than the pressure, as the gradient along $L$ is steeper than the gradient along $p$.

\newpage

\newpage

\section{Discussion}
\label{sec:discussion}

This study aimed to optimize a Parallel-Plate Avalanche Counter with Optical Readout (O-PPAC) for heavy-ion tracking and imaging, focusing on two key detector parameters: pressure and collimator length. A surrogate model was developed to predict the outputs of a \textsc{GEANT4} simulation based on inputs provided to the simulator. By integrating this model into an optimization loop, we identified the optimal combination of parameters that minimizes reconstruction error. The optimization was performed on a batch of $10^5$ alpha particles, and the optimal parameters for a randomly generated distribution of particles covering an area of 8 $\times$ 8 cm$^2$ were found to be a pressure of 39.03 Torr and a collimator length of 15.11 mm.

As shown in Section~\ref{sec:results}, the optimal collimator length is in agreement with previous results obtained with traditional methods, as reported in Reference~\cite{cortesi18}, which validates the approach used in this study. The pressure value, on the other hand, is relatively high within the specified range, which could be attributed to the higher statistics observed at increased pressures. However, this result cannot be explained purely by the statistics, as this would suggest that the optimal pressure should be the highest value within the imposed limits. We hypothesize that other factors, potentially related to the physical behavior of the detector or the limitations of the simulation, may be influencing the outcome.

To ensure the robustness of the results, a grid of different initial configurations was explored. This analysis confirmed that the minimum value of the loss function is independent of the initial values of pressure and collimator length, suggesting that the identified minimum represents the absolute minimum of the reconstruction error. Furthermore, a 3D representation of the optimization curves was used to illustrate the loss function and the difference in the gradient along the two axes. A significantly larger gradient was found along the collimator length axis, indicating that this parameter has a greater influence on the loss function compared to pressure.

These findings align with recent efforts to apply DP techniques to particle and nuclear physics detectors. As remarked before, despite the extensive application of DP in other technical fields, its application to particle detectors remains challenging mainly due to the stochasticity of quantum processes. However, the development of modular differentiable pipelines such as TomOpt \cite{tomopt} is gradually making detector and experiment optimization in nuclear and particle physics more feasible. As proof, this study was successfully conducted by adapting TomOpt software to the specific problem of the O-PPAC optimization.

\section{Summary and Conclusions}

Recent advancements in deep learning and computational capabilities have significantly enhanced the ability to optimize complex systems. Differentiable programming and automatic differentiation are at the forefront of this transformation, enabling automated optimization of complex processes. This study demonstrates the potential of applying these techniques to optimize detectors in nuclear and particle physics. By developing a surrogate model, we were able to identify the optimal parameters for an O-PPAC detector designed for heavy-ion tracking and imaging.

The next steps in this research will focus on incorporating the position reconstruction process directly into the differentiable pipeline, as the expressions applied to the photon distributions in order to obtain the reconstructed position are differentiable. This will involve exploring generative models that can directly predict photon distributions instead of the reconstructed position. Future work will also involve optimizing additional parameters, potentially including cost-related factors in the loss function. Furthermore, we plan to extend this research to a more complex system, which incorporates the O-PPAC as a fundamental component. This could include exploring the effects of higher particle rates, different particle types, and other operational conditions that may influence performance.

In conclusion, this study highlights the potential of differentiable programming and machine learning in the design and optimization of particle detectors. The ongoing work and future directions will further refine this approach, expanding its applicability to a wider range of detector designs and experimental setups.

\section*{Funding}
The work of M. Pereira and X. Cid is supported by the Spanish Research State Agency under projects PID2022-139514NA-C33 and PCI2023-145984-2; by the ``María de Maeztu'' grant CEX2023-001318-M, funded by MICIU/AEI /10.13039/501100011033; and by the Xunta de Galicia (CIGUS Network of Research Centres). Pietro Vischia’s work was supported by the ``Ramón y Cajal” program under Project No. RYC2021-033305-I funded by the MCIN MCIN/AEI/10.13039/501100011033 and by the European Union NextGenerationEU/PRTR. 

The views and opinions expressed are solely those of the authors and do not necessarily reflect those of the European Union or the European Commission. Neither the European Union nor the European Commission can be held responsible for them. 

\section*{Data Availability}
The raw data supporting the conclusions of this article will be made available by the authors on request.

\section*{Acknowledgments}
We express our gratitude to Yassid Ayyad for kindly providing the Geant4 simulation of the detector and for his valuable technical assistance in this work.

\end{document}